\documentclass[12pt]{article}
\usepackage[utf8]{inputenc}
\usepackage{amssymb}
\usepackage{amsmath}
\usepackage{amsthm}
\usepackage{mathtools}
\usepackage{geometry}
\usepackage{setspace}
\usepackage{graphicx}
\usepackage{hyperref}
\usepackage{natbib}
\usepackage{diffcoeff}
\usepackage{xcolor}
\usepackage{booktabs}
\usepackage{listings}
\usepackage{enumitem}
\usepackage{booktabs} % For better table formatting
\usepackage{caption}
\usepackage{subcaption}
\usepackage[english]{babel}
\usepackage{float}
\usepackage{url}
\theoremstyle{plain}
\usepackage{algorithm} 
\usepackage{algpseudocode} 
\usepackage{authblk}
\usepackage{lmodern}

\newcommand{\ftau}{f_{T_I}}

\newcommand{\s}{\tilde{s}}
\newcommand{\iotatil}{\tilde{\iota}}
\newcommand{\rtil}{\tilde{r}}

\newcommand{\X}[1]{\xi_{#1}}
\newcommand{\PP}{\mathbb{P}}
\def\bSig\mathbf{\Sigma}

\allowdisplaybreaks
\begin{document}
   \title{\textmd{Marginal Likelihood Inference for Fitting Dynamical
    Survival Analysis Models to Epidemic Count Data}}
\author[1]{Suchismita Roy\thanks{Email: suchismita.roy@duke.edu}}
\author[1]{Alexander A.~Fisher\thanks{Email: alexander.fisher@duke.edu}}
\author[2]{Jason Xu\thanks{Email: jqxu@g.ucla.edu}}

\affil[1]{Department of Statistical Science, Duke University, Durham, North Carolina, USA}
\affil[2]{Department of Biostatistics, University of California Los Angeles, Los Angeles, California, USA}

\date{}
\maketitle
\vspace{-1cm}
\section*{Abstract}
Stochastic compartmental models are prevalent tools for describing disease spread, but inference under these models is challenging for many types of surveillance data when the marginal likelihood function becomes intractable due to missing information.
To address this, we develop a closed-form likelihood 
for discretely observed incidence count data under the dynamical survival analysis (DSA) paradigm. The method approximates the stochastic population-level hazard by a large population limit while retaining a count-valued stochastic model, and leads to survival analytic inferential strategies that are both computationally efficient and flexible to model generalizations.
Through simulation, we show that parameter estimation is competitive with recent exact but computationally expensive likelihood-based methods in partially observed settings. Previous work has shown that the DSA approximation is generalizable, and we show that the inferential developments here also carry over to models featuring individual heterogeneity, such as frailty models. We consider case studies of both Ebola and COVID-19 data on variants of the model, including a network-based epidemic model and a model with distributions over susceptibility, demonstrating its flexibility and practical utility on real, partially observed datasets.
\paragraph{\textbf{Keywords}:}  Continuous-time Markov Chains; Count Data; Dynamical Survival Analysis; Missing Data; Stochastic Epidemic Models.
\vspace{1cm}
\newpage
\section{Introduction}
Mathematical modeling of infectious disease plays an important role in understanding epidemic dynamics and in guiding public health interventions. Compartmental models are a cornerstone of epidemic modeling,  dividing the population into compartments based on disease status, and describing transitions between compartments based on mechanistic parameters. Classically, these models are  defined by ordinary differential equation (ODE) systems that describe how compartment sizes fluctuate over time. The SIR (Susceptible-Infected-Recovered) model \citep{SIR1927} is perhaps the most well known example, and its many variations are still widely used due to conceptual simplicity, interpretability, and flexibility \citep{britton2020mathematical,morozova2021one,wood2026some}. 

\par While deterministic models are simple to analyze, they fail to describe the inherent randomness of disease spread. Stochastic epidemic models (SEMs) \citep{Anderson_Britton} arguably offer a more realistic representation of disease dynamics. Classical likelihood-based methods enable rigorous inference under \textit{complete} epidemic data, but since infection and recovery histories are rarely fully observed in practice, some of these methods become out of reach in the face of partially observed data. For instance, while trajectories from the stochastic SIR model follow a standard continuous-time Markov chain likelihood \citep{guttorp2017stochastic},  we only observe the process discretely in practice, via case counts, removal data, or other partially observed surveillance studies. Inference becomes substantially more difficult in such settings, as the marginal likelihood now entails a high-dimensional integration step over possible trajectories consistent with the observed data. A large body of work is devoted to overcoming this issue, ranging from numerical methods to compute transition probabilities \citep{Ho-pre_work} and approximating the likelihood using diffusion processes \citep{diffusion_approx} to utilizing particle filters \citep{king_pre_work}, approximate Bayesian computation \citep{MCKinley_pre_work}, and data-augmented MCMC \citep{Gibson_pre_DA,Neil_pre_DA,fintzi2017efficient}.
\par This article revisits this problem from the lens of a recent idea called Dynamical Survival Analysis (DSA) \citep{SDA, DSA_book} to enable efficient and straightforward likelihood-based inference in this setting. Described in more detail below, DSA approximates the stochastic population-level hazard in the stochastic SIR model by its large sample limit. Because it only approximates the hazard rate, the approximating process is defined on the same discrete states as the original model, in contrast to diffusion approximations that imply a continuous population \citep{diffusion_approx}. This technique 
not only tends to provide a faithful approximation to the original epidemic model, but is associated with a survival likelihood that is amenable to simple inferential methods. 
\par The focus of this article is on a rigorous accounting of missing data to understand how these ideas apply and perform on realistic partially observed data types. Specifically, we develop a closed-form likelihood
appropriate for count data that exactly marginalizes over unknown events such as infection and removal times.
In turn, the method enables off-the-shelf posterior sampling algorithms, in contrast to existing works for the partially observed setting discussed above which typically require intensive model-based simulation or bespoke MCMC samplers. We find connections to interval censoring methods in survival analysis, and show that our proposed approach not only leads to more efficient inference than existing methods, but is more readily extensible to relaxations of the model assumptions behind SIR and its variants. To demonstrate,  we showcase its efficacy in models featuring  individual-level variation in susceptibility \citep{frailty1, frailty_theory, Frailty_new}, as well as under Poisson random networks instead of  homogeneous mixing \citep{poisson_nwk} in simulation and several case studies.
\section{Background}
Compartmental models of disease divide individuals into non-overlapping subsets according to their disease statuses. Their dynamics are typically described by ordinary differential equations (ODEs): for instance the susceptible-infectious-removed (SIR) model assumes three disease statuses---susceptible ($S$), infectious ($I$), and removed ($R$) \citep{SIR1927}, and
its dynamics follow 
$\frac{dS(t)}{dt} = -\beta S(t)I(t), \quad \frac{dI(t)}{dt} = \beta S(t)I(t) - \gamma I(t).$

The stochastic SIR model replaces these ODEs instead with similar expressions representing Markov \textit{jump probabilities} or instantaneous rates. In both cases, parameters describe the epidemic \textit{mechanistically}: here $\gamma$ is the rate of recovery for an $I$ individual---in the stochastic model, $1/\gamma$ is
 thus the expected time to recover. Similarly,
$\beta$ can be interpreted as the rate of transmission per contact between $S$ and $I$ individuals. Variations may include additional compartments for latency, reinfection, hospitalization, and other factors \citep{SEIR_SIS2, SEIR_SIS1}. In its simplest form, assuming a  constant, homogeneous rate of infection $\beta$ between susceptible and infectious individuals, the instantaneous infection rate given the current state of the epidemic process at time $t > 0$ is described by 
\begin{equation}
 \mathbb{P}\{(S(t+h),I(t+h))-(S(t),I(t))=(-1,1) \}= \frac{\beta}{N} (S(t)I(t))h+o(h)
 \label{SIR_stochastic_model}
\end{equation}
over a small time increment $h>0$, 
and the recovery process dynamics are defined analogously for a recovery rate $\gamma$. 
The overall stochastic epidemic model evolves as a continuous-time Markov chain (CTMC) \citep{guttorp2017stochastic} where the sojourn time between consecutive events is distributed exponentially. 
Stochastic epidemic models (SEM)---unlike their deterministic counterpart---reflect the randomness behind epidemic spread, and are well-suited to answer questions related to \textit{probabilities} of outbreaks, with arguably more interpretable \textit{uncertainty quantification} for estimates and forecasts. 

We will make use of dynamical survival analysis (DSA) to leverage a survival analytic interpretation of the SIR model, resulting in a simpler likelihood that offers a clear route to marginalizing over  missing data. This approach hinges on a large population limit of an individual-level view of the SIR model that generates the equivalent stochastic process  \citep{Sellke}. Sellke's construction is as follows: 
\begin{enumerate}
    \item Each of the $N$ initially susceptible individuals has an independent infection threshold $Q_i$ exponentially distributed with mean 1. If the cumulative disease exposure exceeds the infection threshold of a susceptible individual, that person will become infected. 
    \item Each infected individual exposes disease to each susceptible at rate $\beta/N$ until recovery/removal, which occurs after an exponential infectious period with rate $\gamma$. 
\end{enumerate}
It is not hard to show that the epidemic generated based on these assumptions follows the same distribution as that generated using \eqref{SIR_stochastic_model} \citep
{book}. Note the total amount of exposure to each individual until time $t$ is 
\begin{equation}
    \Lambda(t) \coloneqq\frac{\beta}{N} \int_0^t I(s)ds. \label{total_exposure_per_person}
\end{equation}
This is the \textit{cumulative hazard} up to time $t$. 
The probability that a random individual will remain susceptible until time $t$ is therefore
\begin{equation}
    \mathbb{P}(Q_i > \Lambda(t) \mid (I(s))_{s \in [0, t]}) = \exp\left( - \frac{\beta}{N} \int_0^t I(s)ds \right). \label{Q_great_Lambda}
\end{equation}   
In this way, Sellke's construction views the SIR model as an individual-level or agent-based process, which will be advantageous for modeling heterogeneity and other features of an epidemic. On the other hand, we can recover the population-based model via lumpability of the process  \citep{lump2}. To see this, we define a process for the $i^{th}$ individual
\begin{align*}
    S_i(t) = \begin{cases}
        1, \text{      if $i^{th}$ individual is susceptible at time $t$.}\\
        0, \text{    otherwise.}
    \end{cases}
\end{align*}
Similarly define the processes $I_i(t)$ and $R_i(t)$ indicating whether person $i$ is infected or recovered respectively at time $t$ so that 
$S_i(t) + I_i(t) + R_i(t) = 1.$  
Summing and by the superposition property of Poisson processes, we arrive at the population-level process $\{S(t), I(t), R(t)\}$,
\begin{equation}
    S(t) = \sum\limits_{i = 1}^{N+M} S_i(t),  I(t) = \sum\limits_{i = 1}^{N+M} I_i(t), \text{    and    }  R(t) = \sum\limits_{i = 1}^{N+M} R_i(t).\label{lumped}
\end{equation}
This yields a useful equivalent representation with unit rate Poisson processes $\mathcal{Y}, \mathcal{Z}$ 
\begin{align*}
    S(t) &= S(0) - \mathcal{Y}\left(\int_0^t\frac{\beta}{n}S(s)I(s)ds\right)\\
     I(t) &= I(0) + \mathcal{Y}\left(\int_0^t\frac{\beta}{n}S(s)I(s)ds\right) - \mathcal{Z}\left(\int_0^t \gamma I(s) ds\right)\\
     R(t) &= \mathcal{Z}\left(\int_0^t \gamma I(s) ds\right).
\end{align*}
The key idea behind DSA is to replace the cumulative stochastic hazard in this construction by its large-sample limit. 
Using the Functional Law of Large Numbers and taking $N \to \infty$ and $M/N \to \rho \in (0, 1)$, \cite{SDA} and \cite{FLLN} showed that the scaled process $(\frac{S(t)}{N}, \frac{I(t)}{N}, \frac{R(t)}{N})$ converges to the solution of the system of ODEs,
\begin{equation}
    \dot{s}_t =  - \beta s_t \iota_t, \dot{\iota}_t = \beta s_t \iota_t - \gamma \iota_t \text{     and   } \dot{r}_t = \gamma \iota_t,\label{final_eqns}
\end{equation}
where,  $  s_t = \lim\limits_{N \to \infty} \frac{S(t)}{N}, \iota_t = \lim\limits_{N \to \infty} \frac{I(t)}{N} \text{     and   } r_t = \lim\limits_{N \to \infty} \frac{R(t)}{N}$
with initial conditions  $s_0 = 1, \iota_0 = \rho, r_0 = 0$.
In other words, $s_t, \iota_t$ are asymptotic proportions of susceptible and infected proportions. The set of ODEs \eqref{final_eqns} can be rewritten as,
\begin{align}
    s_t &= \, \exp\left(-\beta \int_0^t \iota_u du \right)\, \, =  \, \exp(-R_0 r_t),\nonumber\\
    \iota_t &= \rho e^{-\gamma t} - \int_0^t \dot{s}_ue^{-\gamma(t-u)}du, \qquad r_t = \gamma\int_0^t \iota_u du,\label{rewritten}
\end{align}
where $R_0 = \frac{\beta}{\gamma}$ is the basic reproduction number. A key advantage of DSA, detailed below, is that it yields a closed-form density for the times to infection.
\section{Dynamical Survival Likelihoods}
\subsection{Likelihood with Complete Infection and Recovery Times}
To motivate the survival likelihood perspective, observe that the threshold from Sellke's construction \eqref{Q_great_Lambda} is equivalent to an individual's time to infection, inviting a survival analytic view. That is, if $T_I^{(i)}$ is a random variable denoting the time to infection of the $i^{th}$ susceptible, we have $\mathbb{P}(T_I^{(i)} > t \mid (I(s)_{s \in [0, t]} ) = \mathbb{P}(Q_i > \Lambda(t) \mid  (I(s)_{s \in [0, t]})  $. Making use of the relationship between agent-based and population-level formulations of the model in \eqref{rewritten},  \cite{SDA} extend this observation to view  $s$ derived from the ODE limit \eqref{final_eqns} as the survival function for time to infection of a \textit{randomly chosen} susceptible. Denoting this time by $T_I$, the corresponding survival function is then given by
\begin{equation}
    \PP(T_I > t) = s_t = \exp(-\mathcal{R}_0 r_t).\label{survival_function_limiting}
\end{equation}
This view allows us to also interpret $R_0 r_t = \beta \int_0^t \iota_s ds$ as a cumulative hazard function for $T_I$, with hazard rate $\beta \iota_t$ (c.f. \eqref{rewritten}). 
\par From this point of departure, we may next derive the density of $T_I$. Since we do not assume all individuals  become infected by the end of the observation period, the survival function does not converge to $1$ and is therefore an improper random variable. This issue is avoided by considering the conditional density: for $T_I < \infty$, the cumulative distribution function and probability density of $T_I$ are given by
\begin{align}
    F_{T_I}(t_I) &= \frac{1 - s_{t_I} }{\tau}, \qquad
    \ftau(t_I) = - \frac{\dot{s}_{t_I} }{\tau} \label{f_tau},
\end{align}
respectively \citep{SDA, DSA_book}. Here $\tau$ is the probability that a susceptible individual ever becomes infected, and satisfies  
\begin{equation}
    1 - \tau = s_\infty = \exp(- R_0 r_\infty) = \exp(- R_0 (\rho + \tau)).\label{tau}
\end{equation}
For an epidemic observed until time $T$, it follows that
\begin{align}
     \ftau (t; T) &= - \frac{\dot{s}_t}{1 - s_T}, \qquad  F_{T_{I}}(t; T) = \frac{1 - s_t}{1 - s_T}.\label{F_tau_T}
\end{align}
Finally, let $T_R$ denote the recovery time for a randomly selected infected person, and let the infectious period $T_R - T_I$ follow an exponential distribution independent of $T_I$, though this assumption can be relaxed. By convolution of $T_I$ and $T_R - T_I$, the density of $T_R$ is given by
\begin{equation}
    f_{T_R}(t) = \int_0^t \ftau(u)\gamma e^{-\gamma(t-u)}du =  \frac{\gamma}{\tau}\left(\iota_t - \rho \exp(-\gamma t)\right).\label{g_tau}
\end{equation}
Equipped with these explicit densities, we will derive marginal versions of the likelihood that are appropriate for the granularity or availability of the data. \cite{SDA} show that in a completely observed scenario with known infection and recovery times, the likelihood of   $\theta := (\beta, \gamma, \rho)$ for $N$ initial susceptible and $M$ initial infected individuals is
\begin{align}
    &l(\theta \vert \{t_i, w_i\}_{i = 1}^K, \{\epsilon_j\}_{j = 1}^{\tilde{L}}) \nonumber\\
    &= s_T^{N-K} \gamma^{L + \tilde{L}} \exp\Bigg\{-\gamma\left(\sum\limits_{i = 1}^K w_i + \sum\limits_{j = 1}^{\tilde{L}}\epsilon_j + (M - \tilde{L})T \right)\Bigg\} \prod\limits_{i = 1}^K (1 - s_T) \ftau(t_i;T). \label{full_like}
\end{align}
Here,  $t_i$ denotes the exact infection times of each of $K$ individuals who become infected by time $T$, and $w_i$ are the infectious periods or the tails $T - t_i$ for those who recover after $T$. 
$L$ and  $\tilde{L}$ denote the number of initial susceptible and infected individuals, respectively, who recover by time $T$, and $\epsilon_j$ denotes the infectious periods for each of the latter. 
This expression is simply a product of exponentials and allows for straightforward parameter estimation, but we rarely have access to complete data $t_i, w_i, \epsilon_j$ in real-world settings. The next section focuses on the more typical partially observed setting. 
\subsection{Development of Marginal Count Likelihood}
To derive marginal expressions for missing data typical of observational studies, we focus our exposition on two such scenarios. First, we discuss the case when the data contain exact infection times but no information on recoveries. We then consider the more realistic scenario when these infection times are also unavailable, instead summarized by infection counts at some discrete schedule (i.e., daily or weekly reports).  A key observation is that under the DSA approximation, time to infection and time to recovery decouple at the individual level, leading to considerable simplifications in deriving likelihoods marginalized over the missing information. These will translate downstream to efficient and elegant forms that enable user-friendly inference.
\par To illustrate, assume we do not have any recovery or removal information in the data, but observe the exact infection times of $K$ individuals given by $t_1, t_2,\dots, t_K$. Under DSA, the infection times of the sample of $K$ individuals follow $\ftau$ independently, so that for an epidemic observed up until $T$, the likelihood of $\theta$ is
\begin{align}
    l(\theta \vert {t}_{i = 1}^K) = \prod_{i = 1}^K \ftau(t_i;T)
   = \prod_{i = 1}^K  - \frac{\dot{s}_{t_i}}{1 - s_T}
    = \prod_{i = 1}^K  \frac{\beta s_{t_i} \iota_{t_i}}{1 - s_T}. \label{only_infec_like}
\end{align}
Importantly, we see that this likelihood does not depend on $N$, the number of initial susceptibles, which is often unknown. If we were to condition on knowledge of $N$, the likelihood becomes 
\begin{equation}
     l(\theta \vert {t}_{i = 1}^K, N) =  s_T^{N - K} \times \prod_{i = 1}^K (1 - s_T) \ftau(t_i;T)
   = s_T^{N - K} \prod_{i = 1}^K \beta s_{t_i} \iota_{t_i} . \label{s_T_likelihood}
\end{equation}
It is not hard to show that we may also arrive at \eqref{s_T_likelihood}  by marginalizing the complete-data likelihood \eqref{full_like} over all unobserved recovery times.

Exact infection times, however, are rarely available at a granular level in surveillance data.  Suppose now instead we only collect data at discrete observation times $\{\X{0}, \X{1}, \dots, \X{P}\}$ with $\X{0} = 0, \X{P} = T$. We only observe the infection counts $Y_j$ over each interval $(\X{j-1}, \X{j})$, for $j = 1, 2, \dots, P$, and denote the total number of infections before time $T$ by  $K = \sum\limits_{j = 1}^P Y_j.$

To derive the likelihood, we define, $\{t_{j1}, t_{j2}, \dots, t_{jY_j}\}$ as the unobserved infection time of the $Y_j$ individuals in the $j^{th}$ interval $(\X{j-1}, \X{j})$, for $\{j = 1, 2, \dots, P\}$. 
\eqref{s_T_likelihood} gives the density of $\mathbf{t}$, where, $\mathbf{t} = (t_{11}, t_{12}, \dots, t_{1Y_1}, \dots, t_{P1}, \dots, t_{PY_P})$. We will consider a variable transformation,
$\mathbf{t} \rightarrow \mathbf{u},$
where, $\mathbf{u} = (u_{11}, u_{12}, \dots, u_{1Y_1}, \dots, u_{P1}, \dots, u_{PY_P})$,
and the variables are related by the transformation,
 \begin{equation}
   u_{jl} = \frac{s_{t_{jl}} - s_{\X{j-1}}}{s_{\X{j}} - s_{\X{j-1}}}, \text{  for {$j = 1,2,\dots,Y_p, p = 1, 2, \dots, P$,}} \label{u_t_relation}
\end{equation}
Note that since  $\X{j-1} < t_{jl} < \X{j}$ and $s(\cdot)$ is a non-increasing function of time, we have $0<u_{jl}<1$. To derive the density of $\mathbf{u}$, we first calculate the Jacobian. Since, $u_{jl}$ is independent of $t_{j'l'}$, for $j \neq j'$ and $l \neq l'$, the Jacobian is 
\begin{align*}
    J &= \text{det}\left(\text{Diag}\left(\frac{dt_{11}}{du_{11}}, \frac{dt_{12}}{du_{12}}, \dots, \frac{dt_{PP}}{du_{pp}}\right) \right) \\
    &= \prod_{j = 1}^P \prod_{l = 1}^{Y_j} \frac{dt_{jl}}{du_{jl}} = \prod_{j = 1}^P \prod_{l = 1}^{Y_j} \frac{1}{\frac{d u_{jl}}{d t_{jl}}} =  \prod_{j = 1}^P \prod_{l = 1}^{Y_j} \frac{s_{\X{j-1}} - s_{\X{j} } }{\beta s_{t_{jl}} \iota_{t_{jl}}}
\end{align*}   
\citep{rohatgi}. Therefore, denoting the change of variables $t_{jl} = g(u_{jl})$ and substituting into \eqref{s_T_likelihood} reveals the joint density of $\mathbf{u}$ and the incidence count data,
\begin{align*}
f( \mathbf{u}, \{Y_j\}_{j = 1}^P \mid  \theta ) &= s_T^{N - K} \prod_{j = 1}^P \prod_{l = 1}^{Y_j}  \beta s_{g(u_{jl})} \iota_{g(u_{jl})}\mathbb{I}(\X{j-1} < g(u_{jl}) < \X{j})\prod_{j = 1}^P \prod_{l = 1}^{Y_j} \frac{s_{\X{j-1}} - s_{\X{j}} }{\beta s_{g(u_{jl})} \iota_{g(u_{jl})}}\\
&= \underbrace{s_T^{N-K}\prod_{j = 1}^P (s_{\X{j-1}} - s_{\X{j}})^{Y_j}}_{f(\{Y_j\}_{j = 1}^P \mid \theta )}  \times \underbrace{ \prod_{j = 1}^P \prod_{l = 1}^{Y_j} \mathbb{I}(0 < u_{jl} < 1)}_{f(\mathbf{u} \mid \{Y_j\}_{j = 1}^P, \theta )}.
\end{align*}
Hence, each $u_{jl}$ follows a $\text{Unif}(0,1)$ distribution, whose density integrates to one. Remarkably, integrating out the $u_{jl}$ contributes only a factor of 1 and leaves no additional terms. As a result, the marginal likelihood simplifies considerably to
\begin{equation}
     l(\theta \mid \{Y_j\}_{j = 1}^P) = s_{T}^{N-K} \prod\limits_{j = 1}^P (s_{\X{j-1} } - s_{\X{j}})^{Y_j}.\label{ct_like}
\end{equation}
We refer to this quantity as the \textit{marginal} DSA count likelihood, and emphasize that the marginalization over missing events here is the same as the usually difficult integration step described in the introduction. Typically, this must be done using intensive conditional simulation, using filtering approaches or data-augmented MCMC as overviewed earlier.
\paragraph{Connection to interval-censoring}
Upon deriving the above, we found that the expression invites a connection to an interval-censoring view. To see this, introduce  indicator variables
\begin{equation*}
    \delta_{jl} = \begin{cases}
        1, \text{  if $l^{th}$ individual became infected in the interval $(\X{j-1}, \X{j})$}\\
        0, \text{ otherwise,} \qquad l = 1, \dots, Y_j$, \, \,$j = 1, \dots, P.
    \end{cases}
\end{equation*}
Then, the interval-censored likelihood \citep{survival} is given by, 
\begin{align}
    l(\theta \mid \{Y_j\}_{j = 1}^P) &= \mathbb{P}(T_I > T)^{N - K} \prod\limits_{j = 1}^P \prod\limits_{l = 1}^{Y_j}\mathbb{P}\left( \X{j - 1} < T_I < \X{j} \right)^{\delta_{jl}} \nonumber\\
  &=  s_{T}^{N- K} \prod\limits_{j = 1}^P \prod\limits_{l = 1}^{Y_j} (s_{\X{j-1} } - s_{\X{j}})^{\delta_{jl}} = s_{T}^{N-K} \prod\limits_{j = 1}^P (s_{\X{j-1} } - s_{\X{j}})^{Y_j}.\label{ct_like_survival}
\end{align}
An analogous marginal likelihood can be derived from the density of recovery times \eqref{g_tau}, for use with discretely observed removal or death data when these are reported instead of cases.
\paragraph{Comparison to exact inference}
The marginal likelihood \eqref{ct_like} derived from the DSA approximation offer significant computational advantages. For use toward inference, we need only the solution of the ODEs \eqref{final_eqns} at $\{\X{1}, \X{2}, \dots, \X{P}\}$, which are easy to solve numerically. Importantly, this computational cost is independent of the susceptible population size $N$, instead dependent on the number of observation times, so that the method scales well to large outbreaks. In contrast, existing approaches that hinge on the likelihood expressions \eqref{full_like} or \eqref{s_T_likelihood} require a computational budget that increases with $N$ since infection times become denser as the number of susceptibles grows, requiring ODE solutions to be calculated over finer grids. These differences are illustrated empirically in Section \ref{sec:sim}.
\par The following sections show how this methodology, derived in the simple SIR setting for exposition, extends immediately for more complex  models, and how inference is comparable but much faster than computationally intensive gold standards.
\section{Generalizability to Other Compartmental Models}
An advantage of basing inference on the marginal DSA likelihood we derive above is that it implies an inferential strategy that immediately extends to any variant or related compartmental model, as long as the limiting or mean-field dynamics can be specified as a system of ordinary differential equations (ODEs). Fortunately, there is a wealth of mathematical literature in this setting \citep{chowell2017fitting,gumel2021primer,saldana2022modeling}; to motivate this generalizability, we focus on the \textit{frailty model} of individual-level variation and the \textit{Poisson network model} of heterogeneous mixing.
\subsection{Varying susceptibility profiles via frailty}
In the survival analysis literature, deviations from proportional hazards can be modeled by incorporating a \textit{frailty} variable that captures unobserved heterogeneity among individuals \citep{frailty_survival,hougaard1995frailty}, akin to the inclusion of a random effect term. This idea also has utility in compartmental modeling---many models including the standard SIR model assume that all  individuals are equally susceptible, which is often not realistic. Susceptibility may vary due to a variety of factors, including measurable attributes such as age and health conditions, or drivers such as social determinants of health which are difficult to observe \citep{other_health_det1,zelner2022there}. To introduce heterogeneity within the susceptible population, one can introduce a non-negative frailty random variable  \citep{frailty1,talk}. \cite{frailty_theory} derived a closed-form ODE for the SEIR model when the frailty variable follows a Gamma distribution. 
Following their approach, assume that a frailty variable $X$ follows a Gamma distribution with mean $1$ and standard deviation $\nu$ for exposition. Our model extends naturally to accommodate this and allow for joint inference of its parameters with only slight modification. In terms of the model, we modify the Sellke construction by simply now drawing the infection thresholds $Q_i$ from an $\text{Exp}(X_i)$ distribution, instead of identically from $\text{Exp}(1)$. We may understand how each individual's susceptibility is modeled through $X=x$ by analogy to random effects models. Conditional on $X = x$, we may define $S(t, x)$ as the probability that an individual with frailty $x$ remains susceptible at time $t$, and may similarly define $I(t, x), R(t, x)$. 

In terms of the governing equations, we modify by integrating against the frailty distributions, which can be done analytically for exponential families such as the Gamma family. Recalling \eqref{total_exposure_per_person},  a susceptible will remain susceptible until time $t$ with probability
$$ P(Q_i > \Lambda(t) \mid (I(s))_{s \in [0, t]}, x) = S(t, x) = \exp\left( - \frac{x\beta}{N} \int_0^t I(s)ds \right).$$
The infectious quantity is similarly defined here $I(t) = \int I(t, x) dx$; therefore, $$\dot{S}(t, x) = - \exp\left( - \frac{x\beta}{N} \int_0^t I(s)ds \right) \times \frac{x\beta}{N} I(t) = - \frac{x\beta}{N} S(t,x)I(t). $$
For a given individual with frailty variable $x$, we now have the system of ODEs given by
\begin{align}
    \dot{S}(t, x) &= - \frac{x\beta}{N} S(t,x)I(t)  \nonumber\\
    \dot{I}(t, x) &= \frac{x\beta}{N} S(t,x)I(t) - \gamma I(t) \nonumber \\
    \dot{R}(t, x) &= \gamma I(t),\label{indiv_fr}
\end{align}
which mirrors the equations  in \citep{frailty1,frailty_theory} for the SEIR model. For Gamma$(1, \nu)$, integrating against the frailty distribution yields
$ \int xS(t, x) dx = \,S(t)^{1+\nu^2}.$ Thus as $N \to \infty$ and $M/N \to \rho$, the large population behavior is characterized by
\begin{equation}
    \dot{\s}_t =  - \beta \s_t^{1+\nu^2} \iotatil_t, \dot{\iotatil}_t = \beta \s_t^{1+\nu^2} \iotatil_t - \gamma \iotatil_t \text{     and   } \dot{\rtil}_t = \gamma \iotatil_t. \label{final_eqns_frailty}
\end{equation}
We see that in this way, we can immediately use DSA-based likelihoods for inference, simply substituting into the density $\ftau$ given in \eqref{f_tau} which now yields
\begin{equation}
    \tilde{f}_{T_I}(t; T) = \frac{\beta \s_t^{1+\nu^2} \iotatil_t}{1 - \s_T}. \label{fr_f_tau}
\end{equation}
Inference proceeds analogously: the complete likelihood \eqref{full_like} of $\tilde{\theta} = (\beta, \gamma, \rho, \nu)$ is
\begin{align}
    &l(\tilde{\theta} \vert \{t_i, w_i\}_{i = 1}^K, \{\epsilon_j\}_{j = 1}^{\tilde{L}}) \nonumber\\
    &= \s_T^{N-K} \gamma^{L + \tilde{L}} exp\Bigg\{-\gamma\left(\sum\limits_{i = 1}^K w_i + \sum\limits_{j = 1}^{\tilde{L}}\epsilon_j + (M - \tilde{L})T \right)\Bigg\} \prod\limits_{i = 1}^K (1-\tilde{s}_T) \tilde{f}_{T_I}(t_i;T), \label{full_like_fr}
\end{align}
and its marginal form under incidence data modifies the count likelihood \eqref{ct_like} to become
\begin{equation}
    l(\tilde{\theta}\mid \{Y_j\}_{j = 1}^P) = \s_{T}^{N-K} \prod\limits_{j = 1}^P (\s_{\X{j-1} } - \s_{\X{j}})^{Y_j}.\label{ct_like_fr}
\end{equation}
We will use \eqref{ct_like_fr} to estimate the parameters in our simulation study in the next section. 
\subsection{Relaxing well-mixing using a Poisson random network}
To illustrate the model's extensibility, we further consider the SIR model over a Poisson random network studied in \citet{poisson_nwk}. This relaxes the homogeneous mixing assumption behind SIR, which essentially posits that contacts occur over a complete graph, by a random contact network---a CTMC on a random graph $\mathcal{G}(N+M, p)$, where $N+M$ is the size of the graph with degree distribution denoted by $p$. Under the assumption that $p$ follows a Poisson distribution with average degree $\mu$, \cite{poisson_nwk, equivalence_rempala} show that the network SIR model is characterized by the system of equations
\begin{align}
    \dot{S} &= -\tilde{\beta}S(1+\rho -S + \frac{\tilde{\gamma}}{\tilde{\beta}}logS), \label{nwk_s}\\
    \dot{I} &= -\dot{S} -\gamma I, \qquad R = 1+\rho - S - I. \label{nwk-eqn}
\end{align}
under initial conditions $S(0) = 1, I(0) = \rho$, where  $\tilde{\beta} = \mu\beta, \tilde{\gamma} = \beta + \gamma,$ and $\beta, \gamma, \rho, \mu$ are as defined above. As before, the solution curve $S$ satisfying \eqref{nwk_s} enters as the survival function of time to infection, so that the same form for the marginal count likelihood \eqref{ct_like} applies immediately upon substituting $s$  by the solution to the ODEs \eqref{nwk_s},\eqref{nwk-eqn}. 
\section{Simulation Study}\label{sec:sim}
 As previous work on dynamical survival analysis has examined the approximation quality in the completely observed case, our empirical study will focus on assessing how performance using the DSA marginal count likelihood on partially observed data compares to if the complete data were available. Moreover, we compare our approach with exact methods. This section also includes simulation studies in the frailty setting, illustrating the method’s adaptability and effectiveness for model variants.

\subsection{Inference and forward dynamics under DSA}
Before inference or estimation, we begin by first examining the difference in the distribution of outbreaks under the stochastic SIR process and under the approximate dynamics governing DSA. To this end, we simulate $500$ realizations over an observation interval of length $T=10$ from the exact stochastic SIR process (using Sellke's construction). We set the true parameters $\theta = (\beta, \gamma, \rho) = (2, 0.5, 0.05), (2, 1, 0.05), (1.5, 1, 0.05)$ following a study by \cite{SDA}, repeating for populations of size $N = 250, 1000, 10000$.  
\begin{algorithm}
\caption{Pseudocode to generate an infection time $t^*$ from $\ftau$}\label{f_tau_samp}
\textbf{Input:} The parameters $(\beta, \gamma, \rho)$ and the final observation time $T$.\\
\textbf{Steps:}
\begin{algorithmic}[1]
     \State Given the parameters $(\beta, \gamma, \rho)$ solve the ODEs \eqref{final_eqns} for the time interval $(0, T]$ with a fixed small step size and calculate $s_T$.
    \State Generate a random sample $u \sim \text{Unif}(0, 1)$.
    \State Equating $F_{T_I}(t^*; T)$ with $u$, calculate, $s_{t^*} = 1 - u (1 - s_T).$
     \State Construct an interpolating function for the mapping $s_t \mapsto t$ by interpolating the ODE solutions $s_t$ given $t$.
    \State Sample $t^*$ from $s_{t^*}$ via inverse probability sampling using the interpolating function.
\end{algorithmic}
\textbf{Output:} A infection time $t^*$ randomly simulated from the density $\ftau$. 
\end{algorithm}

\begin{figure}
     \centering
     \begin{subfigure}[b]{0.4\textwidth}
         \centering
         \includegraphics[height = 5cm, width = 7cm]{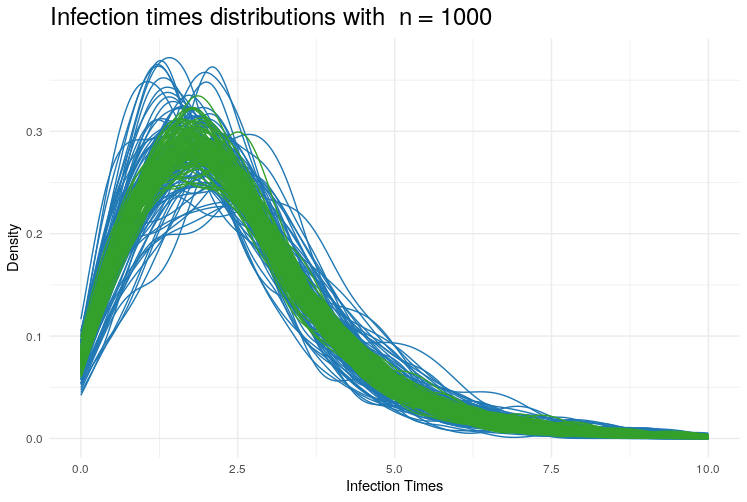}
     \end{subfigure}
     \hspace{1cm}
     \begin{subfigure}[b]{0.4\textwidth}
         \centering
         \includegraphics[height = 5cm, width = 7cm]{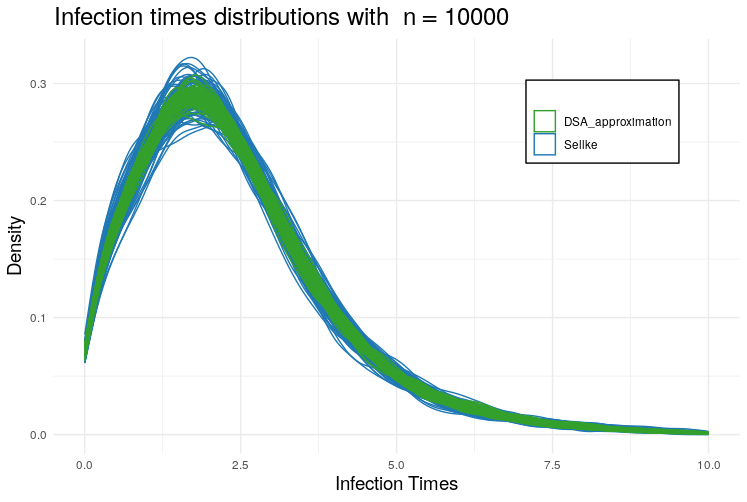}
     \end{subfigure}
     \caption{Comparison of 100 empirical densities of infection time simulated using Sellke construction (in blue) with 100 empirical densities of infection time simulated using DSA approximation (in green).}
     \label{dsa_comp}
\end{figure}

Next, we repeat this study analogously, but replace the exact SIR dynamics by the corresponding approximate dynamics under DSA, detailed in Algorithm \ref{f_tau_samp}. A comparison of sampling $100$ trajectories from the exact SIR and the DSA approximation appears in Figure \ref{dsa_comp}. The plot suggests an empirical difference in the variability reflected under the DSA approximation; we observe the trajectories more closely resemble those under the exact SIR dynamics as $N$ increases, though the gap does not vanish. On the other hand, the mean trends are very similar. 
\par To understand how these differences may translate to differences in estimation, we turn our attention to inference. We compare fitted models under the proposed marginal likelihood to several peer methods: first, we consider parameter estimates using complete information on infection and recovery times using \eqref{full_like}. Next, we compare to a strategy previously applied to case count data by \cite{covidpaper}, \cite{Non-markovian} that imputes the missing infection times as uniform random variables, so that they may conduct inference under the likelihood \eqref{s_T_likelihood}. As a gold standard, we compare to a recent data-augmented MCMC method \citep{Raphael}, that marginalizes over the proper distribution of missing times under the exact SIR model via a bespoke data-augmented sampler.  
\begin{table}[ht]
    \centering
    \footnotesize
    \caption{Parameter Estimation when true parameter is $(\beta, \gamma, \rho) = (2, 0.5, 0.05)$. M1: using exact infection and recovery times; M2: using uniformly approximated infection times; M3: count likelihood; M4: PDSIR. }
    \begin{tabular}{|l|l|cccc|cccc|cccc|}
        \toprule
         & & \multicolumn{4}{|c|}{N = 250} & \multicolumn{4}{|c|}{N = 1000 } & \multicolumn{4}{|c|}{N = 10000} \\
        \cmidrule(lr){3-6} \cmidrule(lr){7-10} \cmidrule(lr){11-14}
        & & avg & sd & 95\% cvg & ESS/s & avg & sd & 95\% cvg & ESS/s & avg & sd & 95\% cvg & ESS/s \\
        \midrule
        \(\beta\) & M1 & 2.06 & 0.29 & 0.82 & 74.51 & 1.99 & 0.07 & 0.85 & 21.84 & 2.00 & 0.02 & 0.83 & 3.38 \\
        & M2 & 1.62 & 0.22 & 0.58 & 23.56 & 1.71 & 0.11 & 0.30 & 10.83 & 1.73 & 0.04 & 0.00 & 1.77 \\
        & M3 & 1.95 & 0.25 & 0.83 & 45.96 & 1.98 & 0.12 & 0.84 & 53.15 & 2.00 & 0.04 & 0.84 & 104.37 \\
        & M4 & 1.98 & 0.24 & 0.94 & 2.78 & 1.99 & 0.12 & 0.94 & 3.91 & 2.00 & 0.04 & 0.92 & 0.71 \\
        \midrule
        \(\gamma\) & M1 & 0.80 & 0.26 & 0.93 & 96.28 & 0.50 & 0.02 & 0.91 & 28.44 & 0.50 & 0.01 & 0.93 & 4.43 \\
        & M2 & 0.39 & 0.09 & 0.75 & 26.55 & 0.42 & 0.05 & 0.59 & 12.10 & 0.43 & 0.02 & 0.01 & 1.83 \\
        & M3 & 0.48 & 0.11 & 0.89 & 51.86 & 0.49 & 0.05 & 0.88 & 59.37 & 0.50 & 0.02 & 0.91 & 116.38 \\
        & M4 & 0.48 & 0.11 & 0.93 & 1.02 & 0.50 & 0.05 & 0.93 & 1.30 & 0.50 & 0.02 & 0.92 & 0.22 \\
        \midrule
        \(\rho\) & M1 & 0.11 & 0.05 & 0.70 & 77.18 & 0.05 & 0.01 & 0.74 & 22.85 & 0.05 & 0.00 & 0.69 & 3.55 \\
        & M2 & 0.10 & 0.03 & 0.54 & 24.38 & 0.08 & 0.01 & 0.37 & 11.62 & 0.07 & 0.00 & 0.00 & 1.77 \\
        & M3 & 0.06 & 0.02 & 0.79 & 47.29 & 0.05 & 0.01 & 0.78 & 57.12 & 0.05 & 0.00 & 0.77 & 113.03 \\
        \bottomrule
    \end{tabular}
    \label{tablep1}
\end{table}
\par  We take $\operatorname{Gamma}(0.1, 0.1)$ as prior for all parameters $\beta, \gamma, \rho$, with the proportion $\rho$ bounded above by $1$. For DSA-based methods, we draw $10^5$ posterior samples using Hamiltonian Monte Carlo via Stan \citep{Neal_HMC,Stan}, 
discarding the first $50\%$ as burn-in. For the exact data augmented sampler, we tune the proportion of missing data to be updated in each Metropolis step to achieve $20-25\%$ acceptance rates.  We considered taking the proportion of updating augmented data (denoted by $\rho$ in the paper) as $(0.4, 0.08, 0.01)$ for three different $N$ of first parameter combination and $(0.4, 0.15, 0.02)$ and $(0.4, 0.35, 0.07)$  respectively for other two set of parameters. The results are summarized in Tables \ref{tablep1}, \ref{tablep2}, \ref{tablep3}, which display posterior means and standard deviations, along with frequentist coverage of the $95\%$ credible intervals and average effective sample size per second.  We see that overall, the posterior mean is close to the true parameters across settings, with the exception of inference under the DSA model when missing event times are imputed uniformly. This observed bias does not vanish as $N$ increases, while all other methods' performance improves as expected. This highlights the need to properly account for the distribution of missing information, which is evidently highly structured and not missing at random in the SIR model.  
\begin{table}[ht]
    \centering
    \footnotesize
    \caption{Parameter Estimation when true parameter is $(\beta, \gamma, \rho) = (2, 1, 0.05)$.
    M1: using exact infection and recovery times; M2: using uniformly approximated infection times; M3: count likelihood; M4: PDSIR.}
    \begin{tabular}{|l|l|cccc|cccc|cccc|}
        \toprule
         & & \multicolumn{4}{|c|}{N = 250} & \multicolumn{4}{|c|}{N = 1000 } & \multicolumn{4}{|c|}{N = 10000} \\
        \cmidrule(lr){3-6} \cmidrule(lr){7-10} \cmidrule(lr){11-14}
        & & avg & sd & 95\% cvg & ESS/s & avg & sd & 95\% cvg & ESS/s & avg & sd & 95\% cvg & ESS/s \\
        \midrule
        \(\beta\) & M1 & 2.02 & 0.21 & 0.74 & 85.28 & 1.99 & 0.07 & 0.75 & 53.40 & 2.00 & 0.02 & 0.77 & 3.14 \\
        & M2 & 1.77 & 0.25 & 0.63 & 28.16 & 1.87 & 0.12 & 0.61 & 9.59 & 1.86 & 0.04 & 0.22 & 2.09 \\
        & M3 & 1.99 & 0.27 & 0.70 & 44.40 & 2.02 & 0.13 & 0.65 & 83.33 & 2.00 & 0.04 & 0.63 & 77.99 \\
        & M4 & 2.03 & 0.37 & 0.95 & 2.07 & 2.02 & 0.18 & 0.94 & 1.08 & 2.00 & 0.06 & 0.93 & 0.80 \\
        \midrule
        \(\gamma\) & M1 & 1.12 & 0.16 & 0.82 & 100.02 & 1.00 & 0.03 & 0.82 & 62.81 & 1.00 & 0.01 & 0.83 & 3.69 \\
        & M2 & 0.90 & 0.14 & 0.65 & 30.10 & 0.94 & 0.07 & 0.65 & 10.22 & 0.94 & 0.02 & 0.27 & 2.23 \\
        & M3 & 1.01 & 0.15 & 0.69 & 47.45 & 1.02 & 0.07 & 0.70 & 83.22 & 1.00 & 0.02 & 0.63 & 83.10 \\
        & M4 & 1.02 & 0.22 & 0.96 & 1.66 & 1.01 & 0.11 & 0.94 & 0.84 & 1.00 & 0.03 & 0.94 & 0.59 \\
        \midrule
        \(\rho\) & M1 & 0.08 & 0.03 & 0.59 & 94.61 & 0.05 & 0.01 & 0.61 & 59.71 & 0.05 & 0.00 & 0.64 & 3.51 \\
        & M2 & 0.09 & 0.03 & 0.64 & 27.86 & 0.06 & 0.01 & 0.65 & 10.13 & 0.06 & 0.00 & 0.22 & 2.23 \\
        & M3 & 0.07 & 0.03 & 0.66 & 44.21 & 0.05 & 0.01 & 0.64 & 83.64 & 0.05 & 0.00 & 0.68 & 83.34 \\
        \bottomrule
    \end{tabular}
    \label{tablep2}
\end{table}
We see that credible interval coverage is below the nominal $95\%$ level for methods under the DSA approximation; this may be unsurprising given the reduced variability we observed in Figure \ref{dsa_comp}. Notably, the coverage deteriorates as $N$ increases when imputing event times uniformly, while the DSA-based methods retain reasonable coverage; this bias is ameliorated for outbreaks with small $R_0 = \beta/\gamma$. To confirm that this undercoverage is a result of the model approximation and not the inferential methodology, we also sample from the posterior of the model when data are simulated from the DSA approximation as described in \ref{f_tau_samp}. Table \ref{table:result_DSA_approx_data} shows that in this case, the nominal coverage is achieved.
\begin{table}[ht]
    \centering
    \caption{Parameter Estimation when true parameter is $(\beta, \gamma, \rho) = (1.5, 1, 0.05)$. M1: using exact infection and recovery times; M2: using uniformly approximated infection times; M3: count likelihood; M4: PDSIR.}
    \footnotesize
    \begin{tabular}{|l|l|cccc|cccc|cccc|}
        \toprule
         & & \multicolumn{4}{|c|}{N = 250 } & \multicolumn{4}{|c|}{N = 1000} & \multicolumn{4}{|c|}{N = 10000} \\
        \cmidrule(lr){3-6} \cmidrule(lr){7-10} \cmidrule(lr){11-14}
        & & avg & sd & 95\% cvg & ESS/s & avg & sd & 95\% cvg & ESS/s & avg & sd & 95\% cvg & ESS/s \\
        \midrule
        \(\beta\) & M1 & 1.54 & 0.18 & 0.81 & 86.93 & 1.50 & 0.06 & 0.83 & 59.63 & 1.50 & 0.02 & 0.83 & 1.99 \\
        & M2 & 1.25 & 0.31 & 0.56 & 16.17 & 1.41 & 0.16 & 0.57 & 23.72 & 1.44 & 0.05 & 0.52 & 1.88 \\
        & M3 & 1.50 & 0.35 & 0.55 & 83.90 & 1.47 & 0.16 & 0.57 & 37.01 & 1.51 & 0.05 & 0.61 & 42.92 \\
        & M4 & 1.52 & 0.37 & 0.94 & 1.65 & 1.53 & 0.20 & 0.94 & 0.88 & 1.50 & 0.06 & 0.94 & 0.23 \\
        \midrule
        \(\gamma\) & M1 & 1.09 & 0.13 & 0.84 & 95.05 & 1.01 & 0.04 & 0.86 & 64.87 & 1.00 & 0.01 & 0.85 & 2.17 \\
        & M2 & 0.86 & 0.21 & 0.58 & 16.83 & 0.95 & 0.10 & 0.58 & 24.27 & 0.96 & 0.03 & 0.50 & 1.92 \\
        & M3 & 1.02 & 0.24 & 0.54 & 86.52 & 0.98 & 0.11 & 0.60 & 37.85 & 1.01 & 0.03 & 0.60 & 43.79 \\
        & M4 & 1.04 & 0.28 & 0.93 & 1.43 & 1.02 & 0.15 & 0.93 & 0.77 & 1.00 & 0.05 & 0.94 & 0.20 \\
        \midrule
        \(\rho\) & M1 & 0.07 & 0.03 & 0.47 & 97.49 & 0.05 & 0.01 & 0.59 & 67.67 & 0.05 & 0.00 & 0.55 & 2.26 \\
        & M2 & 0.16 & 0.10 & 0.65 & 15.63 & 0.07 & 0.02 & 0.68 & 23.51 & 0.06 & 0.00 & 0.58 & 1.94 \\
        & M3 & 0.16 & 0.09 & 0.58 & 75.62 & 0.07 & 0.02 & 0.67 & 36.52 & 0.05 & 0.00 & 0.66 & 44.39 \\
        \bottomrule
    \end{tabular}
    \label{tablep3}
\end{table}

Finally, the DSA-based methods yield orders of magnitude improvements in computational efficiency, owing to the lower autocorrelation under HMC sampling. Compared to inference using the complete data, which is unavailable in real data settings, our marginal likelihood-based inference reveals an inflated variance in estimates as we would expect to see when removing complete event times. Importantly, its computational load does not scale with $N$, in contrast to numerical integration of ODE systems requiring smaller step-sizes that scale with $N$ under the complete data likelihood.
\begin{table}[!htbp]
    \centering
    \caption{Parameter Estimation for datasets simulated from DSA approximation using only infection counts}
    \footnotesize
    \begin{tabular}{|l|cccc|cccc|cccc|}
        \toprule
         &  \multicolumn{4}{|c|}{N = 250} & \multicolumn{4}{|c|}{N = 1000} & \multicolumn{4}{|c|}{N = 10000} \\
        \cmidrule(lr){2-5} \cmidrule(lr){6-9} \cmidrule(lr){10-13}
        & Avg & SD & 95\% Cvg & ESS/s & Avg & SD & 95\% Cvg & ESS/s & Avg & SD & 95\% Cvg & ESS/s \\
        \midrule
        $\beta = 2$   & 1.94 & 0.25 & 0.93 & 36.47 & 1.98 & 0.12 & 0.95 & 16.37 & 2.00 & 0.04 & 0.95 & 55.17 \\
        $\gamma = 0.5$ & 0.47 & 0.10 & 0.92 & 41.22 & 0.49 & 0.05 & 0.95 & 18.30 & 0.50 & 0.02 & 0.95 & 61.49 \\
        $\rho = 0.05$ & 0.06 & 0.02 & 0.93 & 37.33 & 0.05 & 0.01 & 0.96 & 17.59 & 0.05 & 0.00 & 0.95 & 59.80 \\
        \midrule 
        $\beta = 2$   & 1.98 & 0.26 & 0.93 & 43.06 & 1.99 & 0.13 & 0.96 & 47.04 & 2.00 & 0.04 & 0.94 & 48.79 \\
        $\gamma = 1$  & 0.99 & 0.14 & 0.94 & 46.07 & 0.99 & 0.07 & 0.95 & 50.14 & 1.00 & 0.02 & 0.95 & 51.93 \\
        $\rho = 0.05$ & 0.06 & 0.02 & 0.94 & 43.15 & 0.05 & 0.01 & 0.96 & 49.73 & 0.05 & 0.00 & 0.94 & 52.19 \\
        \midrule 
        $\beta = 1.5$ & 1.38 & 0.34 & 0.91 & 112.28 & 1.49 & 0.16 & 0.92 & 41.01 & 1.50 & 0.05 & 0.95 & 49.56 \\
        $\gamma = 1$  & 0.93 & 0.22 & 0.91 & 115.51 & 0.99 & 0.10 & 0.92 & 41.86 & 1.00 & 0.03 & 0.94 & 50.54 \\
        $\rho = 0.05$ & 0.09 & 0.07 & 0.91 & 89.17 & 0.05 & 0.01 & 0.93 & 40.86 & 0.05 & 0.00 & 0.94 & 51.26 \\
        \bottomrule
    \end{tabular}
    \label{table:result_DSA_approx_data}
\end{table}
\subsection{Frailty models}
We now move to conduct a simulation study under a frailty  model framework. In this experiment, the data sets are simulated via a modified Sellke construction with $Q_i$ following the order statistics of  $\operatorname{Exponential}(X)$, where $X \sim Gamma(1/\sqrt{\nu}, 1/\sqrt{\nu})$.  This induces a hierarchy of variability through $X$ and introduces an additional parameter $\nu$ describing the frailty distribution, with the rest of the simulation design unchanged. 
\begin{table}[!htbp]
    \centering
    \caption{Parameter estimation for frailty models with only infection counts, 
    for three parameter sets: 
    $(\beta, \gamma, \rho, \nu) = (2, 0.5, 0.05, 0.1); \, (2, 1, 0.05, 1); \, (1.5, 1, 0.05, 0.5)$.}
    \label{frailty_result}
    \footnotesize
    \begin{tabular}{|l|cccc|cccc|cccc|}
        \toprule
         & \multicolumn{4}{c|}{N = 250} & \multicolumn{4}{c|}{N = 1000} & \multicolumn{4}{c|}{N = 10000} \\
        \cmidrule(lr){2-5} \cmidrule(lr){6-9} \cmidrule(lr){10-13}
         & Avg & SD & 95\% Cvg & ESS/s & Avg & SD & 95\% Cvg & ESS/s & Avg & SD & 95\% Cvg & ESS/s \\
        \midrule
        $\beta=2$   & 1.93 & 0.26 & 0.82 & 22.32 & 1.99 & 0.13 & 0.83 & 49.48 & 2.00 & 0.04 & 0.86 & 30.10 \\
        $\gamma=0.5$ & 0.35 & 0.14 & 0.78 & 17.74 & 0.42 & 0.09 & 0.82 & 19.23 & 0.48 & 0.03 & 0.94 & 11.27 \\
        $\rho=0.05$  & 0.06 & 0.02 & 0.82 & 22.75 & 0.05 & 0.01 & 0.83 & 43.78 & 0.05 & 0.00 & 0.80 & 24.03 \\
        $\nu=0.1$   & 0.29 & 0.17 & 0.98 & 17.73 & 0.23 & 0.13 & 0.97 & 19.87 & 0.13 & 0.07 & 0.97 & 32.71 \\
        \midrule
        $\beta=2$   & 1.91 & 0.52 & 0.73 & 52.15 & 2.08 & 0.28 & 0.71 & 14.61 & 2.05 & 0.08 & 0.68 & 38.57 \\
        $\gamma=1$  & 1.09 & 0.37 & 0.87 & 53.72 & 1.18 & 0.27 & 0.77 & 10.52 & 1.10 & 0.13 & 0.69 & 34.01 \\
        $\rho=0.05$ & 0.13 & 0.09 & 0.76 & 44.46 & 0.06 & 0.02 & 0.76 & 14.01 & 0.05 & 0.01 & 0.82 & 40.00 \\
        $\nu=1$     & 0.78 & 0.44 & 0.98 & 60.31 & 0.76 & 0.34 & 0.87 & 9.91  & 0.90 & 0.15 & 0.73 & 31.28 \\
        \midrule
        $\beta=1.5$ & 1.58 & 0.35 & 0.72 & 77.23 & 1.47 & 0.21 & 0.68 & 15.30 & 1.52 & 0.09 & 0.68 & 3.66 \\
        $\gamma=1$  & 0.81 & 0.32 & 0.74 & 57.50 & 0.86 & 0.23 & 0.79 & 11.04 & 1.00 & 0.12 & 0.80 & 3.08 \\
        $\rho=0.05$ & 0.05 & 0.02 & 0.88 & 81.13 & 0.07 & 0.03 & 0.72 & 16.27 & 0.05 & 0.00 & 0.64 & 6.16 \\
        $\nu=0.5$   & 0.90 & 0.45 & 0.89 & 66.53 & 0.67 & 0.34 & 0.94 & 11.79 & 0.47 & 0.20 & 0.86 & 3.11 \\
        \bottomrule
    \end{tabular}
\end{table}
We consider three parameter combinations here:  $(\beta, \gamma, \rho, \nu) =$ $ (2, 0.5, 0.05, 0.1),$ $ (2, 1, 0.05, 1),$ $(1.5, 1, 0.05, 0.5)$. For each setting, we again consider three different initial susceptible populations, $N = 250, 1000, 10000$, and simulate $500$ data sets using the modified Sellke construction for each case, keeping only the daily infection counts while ignoring continuous infection times and recovery times. Then, we estimate the parameters using \eqref{ct_like_fr}, where $\tilde{s}_t$ is the solution of the ODEs \eqref{final_eqns_frailty}. Table \ref{frailty_result} summarises the results for all four parameters.
\par We note that mild practical unidentifiability can be observed.
Because the frailty distribution introduces additional parameters, more data can be necessary to successfully estimate all parameters. Nonetheless, we see that moving to a slightly less diffuse $\operatorname{Gamma}(1, 1)$ prior largely remedies the issue. 
Estimation of the key epidemic parameters  $\beta$ and $\gamma$ is successful with similar coverage as the homogeneous model, even when estimating frailty parameters such as $\nu$ becomes more difficult, for instance in smaller populations. Overall, this
simulation confirms the efficacy of the count likelihood for frailty models of heterogeneity, though more informative priors can be helpful in avoiding practical identifiability in richer parameterizations. 
\section{Case Studies}
We now apply our method to study  two data sets---the  2018–2020 Ebola
outbreak in the Democratic Republic of the Congo, and the early COVID-19 pandemic of 2020 in China.
\paragraph{Poisson Network SIR Model for Ebola Outbreak}
In this section, we demonstrate the performance of the count likelihood for the Poisson-network SIR model by analyzing the third wave of the 2018–2020 Ebola outbreak in the Democratic Republic of the Congo (DRC). Two prior analyses analyze these data using DSA: both approaches assume or impute knowledge on all unobserved epidemic times, with \cite{ebola} fitting an SIR model, and \cite{poisson_nwk} revisiting the data under a Poisson random network. Equipped with the marginal likelihood we may revisit this data and directly consider daily case counts. Doing so does not make the strong assumption that exact times of infection and recovery can be informed from the data, which are based on hospitalization records, recorded symptom onsets, or self reported home recovery. A more detailed description of the dataset is available in \cite{ebola} and \cite{poisson_nwk}. 
\par Following \cite{poisson_nwk} we estimate the unknown size $N$ via
$\hat{N} = \frac{K_T}{1 - S_T},$
where $K_T$ is the total number of infected until time $T$ and $S$ is the solution of $\eqref{nwk_s}$. The outbreak is observed over a window of $T = 108$ days, comprising $1068$ infection events. 
We also follow their prior specification with distributions $ \tilde{\beta}, \tilde{\gamma} \sim \operatorname{Gamma}(0.02, 0.02)$ and $\rho $ following a uniform distribution,  with the constraints 
   $\tilde{\beta} \in (0.1, \infty), \tilde{\gamma} \in (0, \tilde{\beta}), \rho \in (0, 0.015).$
 Table \ref{nwk} compares the estimates using the complete infection and recovery time of the paper \cite{poisson_nwk} and our estimates using count likelihood. 
\begin{table}[!htbp]
    \centering
    \caption{Comparing estimates of the parameters of Poisson network model for Ebola data}
    \begin{tabular}{|c|c|c|}
    \toprule
        Parameter & Complete Information & Count likelihood  \\
        \midrule
       $\tilde{\beta}$ & 0.229 (0.209, 0.259) & 0.227(0.155, 0.447) \\
        $\tilde{\gamma}$ & 0.215 (0.197, 0.242) & 0.205 (0.134, 0.426) \\
        $\rho$ & 0.0055 (0.0046, 0.0073) &  0.009 (0.002, 0.015)\\
        $\mu$ & 39.48 (7.93, 93.00) & - \\
        $\tilde{R}_0$ & 1.071 (1.034, 1.109) & 1.121 (1.044, 1.200)\\
        \bottomrule
    \end{tabular}
    \label{nwk}
\end{table}
As inference is now based only on counts with no additional assumption on event times, the posterior credible intervals reflect more uncertainty under our marginal analysis. The results are consistent with previous analyses in that their narrower intervals are contained within our credible intervals, with similar posterior means. These wider intervals may better reflect the data, as it may not be realistic to treat symptom onsets as the exact infection times under a continuous-time model. Our analysis leads to a smaller estimated final size of the outbreak at $3502$, compared to $3773$ in the previous analysis, with overlapping credible intervals. 
\paragraph{COVID-19 in China}
We next apply our proposed methodology to analyze COVID-data from China  \citep{coviddata_citation}. 
We considered incidence data comprising daily infection counts
between January 25 and March 5, 2020, obtained in 53 cities. Mirroring the study in \cite{china_data} to ensure a fair comparison, we also fix $\gamma = 1/6$ and apply analogous preprocessing, which suffice to remove practical identifiability. The population size of the cities, with a minimum of 0.5 million, is sufficiently large compared to the sample size, which is less than 3,000, to justify the asymptotic assumptions behind DSA. 

We fit both the DSA approximation to the stochastic SIR model using the marginal likelihood \eqref{ct_like}, as well as a frailty model variant allowing for hetereogeneity in the susceptible population using the likelihood \eqref{ct_like_fr}.   We sample from the posterior using HMC via Stan under a $\operatorname{Gamma}(0.2, 0.2)$ prior for $\beta, \nu$ and $\operatorname{Uniform}(0, 1)$ prior for $\rho$. To assess the model performance, we overlay the densities of infection times against the raw histograms of daily infection counts from the data. Density plots are derived from the posterior mean estimates of model parameters after running Hamiltonian Monte Carlo on each of the two models; these results are consistent with the findings by \cite{china_data}. 
\begin{figure}
     \centering
     \begin{subfigure}[b]{0.4\textwidth}
         \centering
         \includegraphics[height = 5.5cm, width = 8cm]{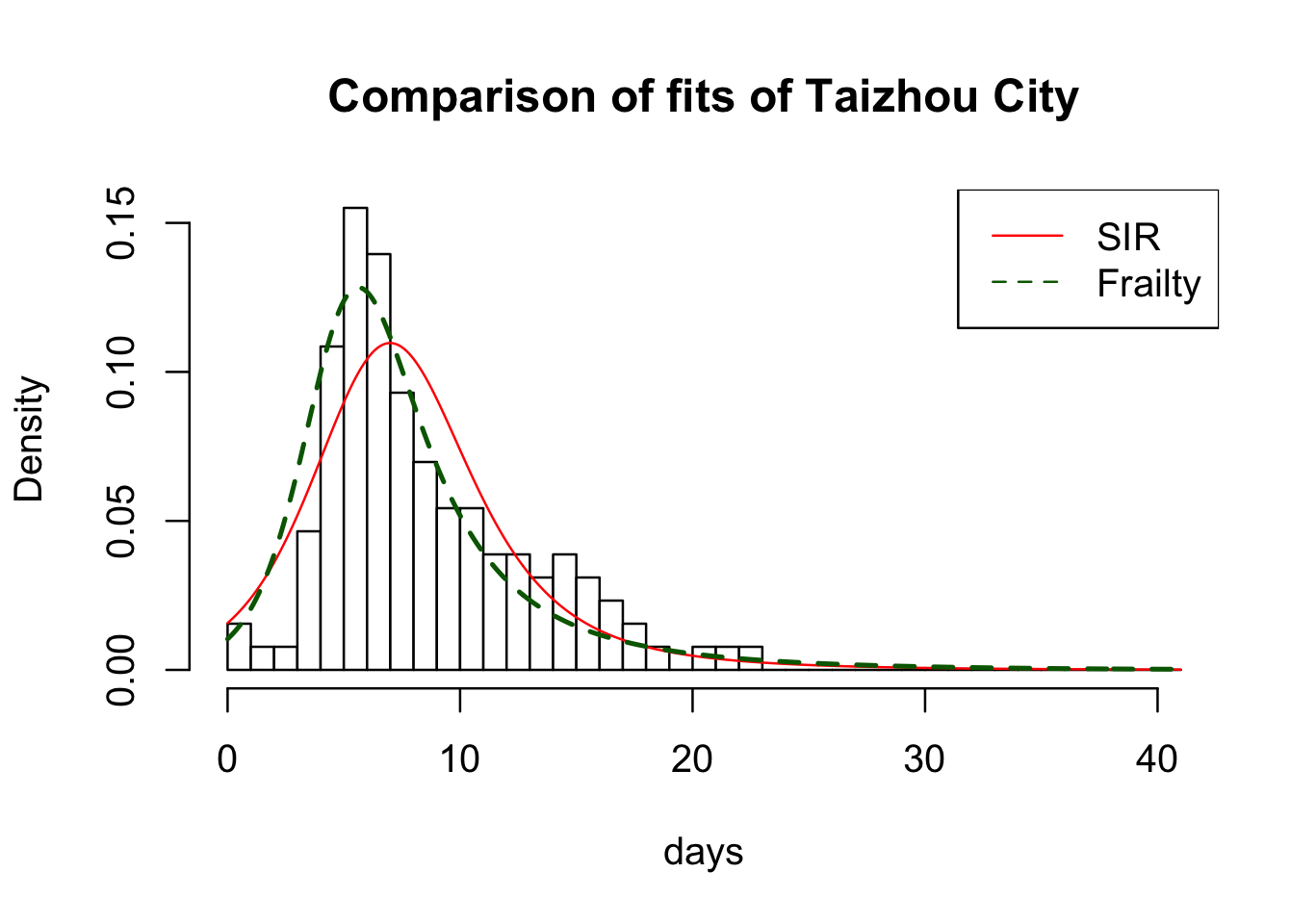}
     \end{subfigure}
     \hspace{1cm}
     \begin{subfigure}[b]{0.4\textwidth}
         \centering
         \includegraphics[height = 5.5cm, width = 8cm]{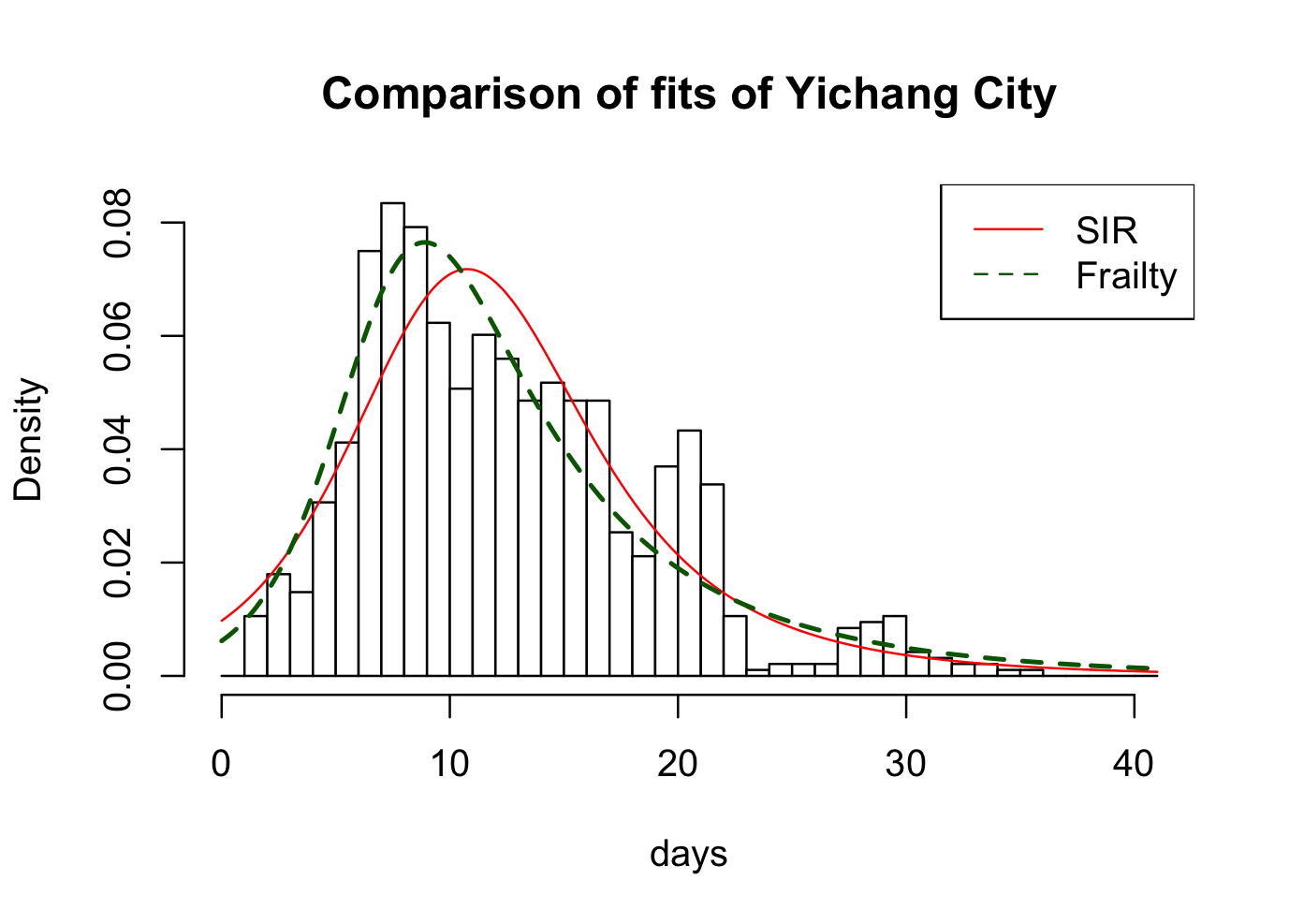}
     \end{subfigure}
     \caption{Histogram of daily infection counts (normalized to area 1), overlaid with the estimated density of infection times for the standard SIR model and the frailty model.}
     \label{Taizhou}
\end{figure}
\par Our analysis suggests that allowing for varying susceptibility leads to an equal or superior fit compared to the standard SIR model. The parameter estimates are largely in agreement, while for a number of cities, the estimated frailty parameter $\nu$ is well separated from zero. In such cases, we find that the model fit improves using the frailty model; Figure \ref{Taizhou} displays these findings in two of these cities. We see that the density plot is better fit to the raw counts under varying susceptibility than in the homogeneous model. Importantly, we observe that the estimated parameters are similar between the standard SIR and the frailty model, but the posterior mean estimates of the basic reproductive number $R_0$ for these cities are not contained within the $95\%$ credible intervals of the standard SIR model fit. 
This suggests that despite the visually similar densities at the population level, differences in estimates of key epidemic parameters may be statistically significant. More details and complete posterior estimates and model checks are available at the Github repository linked in the Supplementary materials.

As the frailty model appears to yield a superior fit, this suggests potential biases in inferred epidemic parameters when differing susceptibility among individuals is overlooked. Both models capture the overall trends and shape of the epidemic to a certain extent. However, in nearly all cities, significant variation remains unexplained. This suggests that more sophisticated models that account for factors such as contact networks, intervention measures, and viral evolution could be explored \citep{fanbu, SEIR_SIS2, David_rasmussen} to provide a more complete understanding of the epidemic dynamics.  
\section{Discussion}
This paper develops a closed-form likelihood toward filling the gap in methodology for fitting stochastic epidemic models under missing data scenarios. We have focused on incidence count data, but the methodology applies analogously for discrete removal data, or when both are available and partially observed. While we have focused on posterior sampling in a Bayesian framing in this article, the insights we derive regarding inference under the DSA approximation and the effects under various methods for accounting for missing data apply to  frequentist likelihood-based frameworks as well. To the best of our knowledge, this is the first closed-form likelihood for count data under DSA approximation, and makes the case that it provides an appealing modeling paradigm so that marginalization and downstream inference is user-friendly even as we consider variations to the model.  
\par Many existing methods for fitting count data are developed in the context of specific epidemic models, but we have seen that the proposed approach to direct marginal likelihood inference is readily extensible to many other compartmental models, as long the large-sample behavior can be described by an ODE system. Though not quite as general as simulation-based ``plug-and-play" methods, the method is significantly lighter computationally, running on the order of seconds on a standard laptop. While our empirical results are encouraging, future work may seek to derive correction factors on the ``missing" variances to further improve coverage properties of credible intervals. It will be fruitful to study connections to related approaches using a tractable stochastic process such as linear noise approximation \citep{fintzi2022linear}.

Similarly, the case study reveals that while the standard SIR model under DSA as well as the frailty modification capture overall epidemic trends well, substantial variability remains. This suggests more realistic models incorporating contact networks, interventions, detailed immune profiles, or viral evolution may further improve model realism and expressiveness. Because some level of practical identifiability may arise with more complex models, careful considerations of the level of prior information or shrinkage may guide practice, and integrating data from multiple sources may be fruitful to this end. Our exposition has focused on Markov models, but the  methodology can be readily extended to non-Markovian settings \cite{Non-markovian} while largely retaining its simplicity and ease of use. As the framework is considerably extensible and modular, we hope these contributions will invite further perspectives at the interface of mathematical modeling of disease dynamics, survival analysis, and stochastic process modeling.

\section*{Supplementary Materials}
The COVID-19 data are accessible at  \url{https://dataverse.harvard.edu/dataset.xhtml?persistentId=doi:10.7910/DVN/MR5IJN}. All other data and a complete implementation with open source code are available at \url{https://github.com/Suchismita-Statistics/DSA.CountData}.

\section*{Acknowledgments}
This research was partially supported by the Center of Excellence for Multiscale Immune Systems Modeling (NIH 1U54-AI191253) and by NSF RAISE grant DMS-2230074.

\setcitestyle{numbers}
\bibliographystyle{apalike}
\bibliography{Reference.bib}

\end{document}